\documentclass{article}
\usepackage{arxiv}
\usepackage[utf8]{inputenc} 
\usepackage[T1]{fontenc}    
\usepackage{hyperref}       
\usepackage{url}            
\usepackage{booktabs}       
\usepackage{nicefrac}       
\usepackage{microtype}      
\usepackage{lipsum}

\usepackage{cite}
\usepackage{amsmath,amssymb,amsfonts}
\usepackage{pifont}
\usepackage{multirow}
\usepackage{algorithmic}
\usepackage{graphicx}
\usepackage{textcomp}
\def\BibTeX{{\rm B\kern-.05em{\sc i\kern-.025em b}\kern-.08em
    T\kern-.1667em\lower.7ex\hbox{E}\kern-.125emX}}
    
\title{LUCE: A Blockchain-based data  sharing  platform  for monitoring  data license  accountability  and  compliance}

\author{
Visara Urovi \\
  Institute of Data Science \\
  Maastricht University\\
  6211 LK Maastricht, The Netherlands \\
  \texttt{v.urovi@maastrichtuniversity.nl} \\
  \And
  Vikas Jaiman \\
  Institute of Data Science \\
  Maastricht University \\
   6211 LK Maastricht, The Netherlands \\
  \texttt{v.jaiman@maastrichtuniversity.nl} \\
  \And
  Arno Angerer \\
  School of Business and Economics \\
  Maastricht University\\
  6211 LK Maastricht, The Netherlands \\
    \texttt{angerer.arno@googlemail.com} \\
   \And
 Michel Dumontier \\
 Institute of Data Science  \\
  Maastricht University\\
  6211 LK Maastricht, The Netherlands \\
    \texttt{michel.dumontier@maastrichtuniversity.nl} \\
}

\begin{document}
\maketitle

\begin{abstract}
Easy access to data is one of the main avenues to accelerate scientific research. As a key element of scientific innovations, data sharing allows the reproduction of results, helps prevent data fabrication, falsification, and misuse. Although the research benefits from data reuse are widely acknowledged, the data collections existing today are still kept in silos. Indeed, monitoring what happens to data once they have been handed to a third party is currently not feasible within the current data-sharing practices. We propose a blockchain-based system to trace data collections, and potentially create a more trustworthy data sharing process. 
In this paper, we present the LUCE (License accoUntability and CompliancE) architecture as a decentralized blockchain-based platform supporting data sharing and reuse. LUCE is designed to provide full transparency on what happens to the data after they are shared with third parties. 
The contributions of this work are: the definition of a generic model and an implementation for decentralized  data sharing accountability and compliance and to incorporates dynamic consent and legal compliance mechanisms. We test the scalability of the platform in a real-time environment where a growing number of users access and reuse different datasets.
Compared to existing data-sharing solutions, LUCE provides transparency over data sharing practices, enables data reuse and supports regulatory requirements. The experimentation shows that the platform can be scaled for a large number of users.
\end{abstract}



\section{Introduction}
\label{sec:introduction}
The scientific community recognizes the importance of data sharing~\cite{OpenData2017, Nelson2009}. As a key element of scientific research, data sharing allows the reproduction of scientific results~\cite{Borgman2012, Nelson2009, OpenData2017}, helps prevent data fabrication and falsification~\cite{Tenopir2011}. Currently, data sharing is often required by funding bodies, publishers~\cite{Borgman2012, OpenData2017} and several EU and US funding initiatives~\cite{Elixir2017,EOSC2018, NCATS2018,DataCommons2018}.

Although the benefits of data sharing and reuse are widely acknowledged, evidence of data sharing practices is limited. Currently, a majority of researchers share their data only directly, that is, from person to person (e.g. by email), and mostly with collaborators, which suggests that trust is an important factor in sharing data. One-third of researchers do not share their data at all and public data sharing occurs through the appendix of research articles, stand-alone publications in data journals, repositories, personal websites, and specific websites~\cite{OpenData2017,Borgman2012}. These ways of data sharing make it hard to find data even if it is shared publicly.

Several factors explain the large gap between the agreed importance of sharing data and today's actual practices~\cite{OpenData2017, Borgman2012, WhitePaper2018, Nelson2009} (i) There is no commonly accepted definition of what data sharing exactly means and which data should be shared. (ii) Researchers lack the expertise, training, infrastructure, and resources, to share their data. (iii) Researchers rarely receive credit for sharing data, partly because how to cite and attribute data is not commonly defined yet. (iv) Privacy concerns, control over what happens to the data, and ethical issues prevent researchers from sharing their data.

One way to deal with these concerns is to use data licensing~\cite{Stodden2009}. Licenses clearly state the data reuse conditions, thereby creating legal clarity for the researchers who reuse the data. Yet, due to a lack of awareness over data licensing, a significant proportion of shared data is not licensed. Even when data are licensed, restrictive licenses are most often chosen due to the difficulty of tracking how data are used once they have been shared~\cite{OpenData2017}. 
The ability to track data reuse is crucial When the data contains information of data subjects. In this context, the General Data Protection Regulation (GDPR)~\cite{GDPR} is an important regulation describing the right to access for individuals. Data subjects are given access rights to access information about their individual data, such as who is using the data and for which purpose. Two other elements of the GDPR are of high significance in the context of data sharing and reuse: the individual's rights to erasure (involving the deletion of individual records) and to rectification (involving the update of individual records).
In general, the lack of traceability and clarity on the secondary use of data is driving the unwillingness to share datasets. Centralized platforms have already been proposed with the aim of facilitating data sharing (Google data~\cite{google}, kaagle~\cite{kaggle}, GitHub~\cite{github}), however, these platforms are controlled by one single authority and do not properly address the traceability and clarity of secondary data use. 
A blockchain-based solution can overcome many of the challenges of data reuse. Blockchain technology provides a methodology for constructing data structures that track data transaction history in an immutable manner. 
A blockchain based data structure is designed to be append only, thus, it traces the data reuse in a verifiable, tamper-proof immutable history of transactions. Defining a blockchain based model for data reuse provides a guarantee of a transparent data exchange process. 
Blockchain enabled data sharing has been proposed for different domains, such as health data exchange (i.e.\cite{ jaiman2020consent, zerka2020blockchain, fan2018medblock, agbo2019blockchain}), the internet of things\cite{fernandez2018review}, energy \cite{andoni2019blockchain, wongthongtham2021blockchain} and many other use cases. In these scenarios, blockchains are prototypes resulting in platform designs that are not generic enough to be used for data-sharing and reuse. Some works focusing on enabling generic data sharing platforms with blockchain technology have been described in~\cite{Neisse2017,Ramachandran2017, Liang2017, Ocean}. These works cover many relevant issues to the data sharing process, such as data subject inclusion~\cite{Neisse2017} and enforcement of regulations via smart contracts~\cite{Ramachandran2017}, however fall short into designing the data sharing process for a transparent data sharing and reuse: including how records are explored and accessed, how they are reused in relation to user consent and how to revoke or update the records based on the requests from data subjects.

To overcome these limitations, we present LUCE, a blockchain-based platform that addresses key challenges in data sharing and reuse. LUCE is a web-based accessible solution for data subjects, data providers, data requesters as well as data authorities. The prototyped model facilitates compliance with data licensing terms and helps researchers to comply with the rights to access, rectification, and erasure aspects of the GDPR. LUCE facilitates data sharing and the verification of what happens thereafter. The contributions of this work are twofold: the definition of a generic decentralized data sharing model for accountability and compliance with data re-use and the definitions of  dynamic consent and legal compliance mechanisms. We implement and test the scalability of the platform in a real-time environment where a growing number of users access and reuse different datasets. We conclude that the solution provides transparency over data sharing practices, enables data reuse and supports regulatory requirements. The experimentation shows that the platform can be scaled for a large number of users.
The remainder of this paper is organized as follows. In section \ref{sec:background}, we introduce the building blocks of our solution, namely blockchain technology, smart contracts, and the EU General Data Protection Regulation. In section \ref{sec:architecture}, we describes the LUCE architecture and the interaction protocol. Thereafter we present the implementation of LUCE in section \ref{sec:implementation} and the evaluation of the platform in section \ref{sec:evaluation}. Further, in section \ref{sec:related}, we follow with the current state of the art for data sharing solutions. We conclude this paper in section \ref{sec:conclusions} with a discussion on current and future work.

\section{Background}\label{sec:background}
In this section, we briefly discuss background work on blockchain, smart contracts, and GDPR. 
\subsection{Blockchain} Blockchain can be thought of as an append-only and distributed transactional database~\cite{nakamoto2008bitcoin, edurekaBlockchain1}. A blockchain network has three main components: (i) the blockchain itself, that is, the file containing the records of all transactions. (ii) the peer-to-peer (P2P) network where the participants interact via the blockchain protocol to transact and update the blockchain. (iii) the consensus mechanism. The records of all transactions in the blockchain network are stored in the blockchain. Transactions are organized in blocks that are linked to each other in chronological order. Each block contains the records of transactions and the identifier of the block preceding it in the chain~\cite{nakamoto2008bitcoin}. 
The records of transactions that are stored in the blockchain are auditable and verifiable but cannot be modified once they have been added. This is achieved through the use of cryptographic hashing~\cite{hashing}. A hash function maps the data of variable length to a fixed-length digest. Any change to the input data results in an unpredictable change in the hash. In a blockchain model, every transaction is added to a data structure (namely called the chain) which includes a hash of the previous transactions. If any of the transactions is changed at any point, the subsequent hash would no longer be valid.
Given that there is no single central authority managing the blockchain, a consensus mechanism is necessary to formally encode rules regarding how transactions are validated and how they are added to the ledger. These transactions are verified and validated by the specific nodes, called \textit{miners} who append new blocks in the blockchain. Mining can be an activity open to all or controlled to specific nodes. The latest model is also known as a permissioned blockchain model. Permissioned blockchains have seen some adoption in several business domains however they have shown low adoption from individuals \cite{helliar2020permissionless}. By design, they limit decentralisation to a consortium of partners and that is why we focus on open blockchain models (also known as permissionless). In open blockchains, miners typically add blocks by solving a computationally expensive cryptographic puzzle. Whoever solves it first, receives a monetary reward. The miner who creates a new candidate block broadcast it to the network. Using the consensus protocol, the nodes verify and validate the candidate block which is afterwards added to the blockchain. Cryptographic puzzles have been proven to be expensive for the scalability and for the energy consumption of blockchain systems \cite{chauhan2018blockchain}, therefore alternative protocols to cryptographic puzzles (also known as proof of work models) are also available (such as proof-of-stake models \cite{saleh2020blockchain}).  

\noindent\textbf{Smart Contract:} 
A smart contract is a computer code that runs on the top of a blockchain network and that contains rules - rights and obligations - defining the interaction between the parties to the smart contract~\cite{smartContract}. When all parties meet pre-defined conditions, the smart contract automatically enforces the agreement between them. Transactions with the smart contracts are recorded within the blockchain on the top of which it is running. Smart contracts allow for several parties who do not especially trust each other to transact with each other without a trusted third party.
Smart contracts are used for simple economic transactions but also for other, more complex, purposes, such as registering ownership, intellectual rights, or data-sharing practices. 
\subsection{Data Sharing Policies} The EU General Data Protection Regulation (GDPR) came into effect in May 2018 to extend the requirements of organizations about collecting and processing personal data of EU residents~\cite{GDPR}. The GDPR applies to EU and non-EU organizations that collect and process EU residents’ data. 
The main actors involved in the GDPR are: the \textit{data subject} - an identified or identifiable natural person whose data are contained in the dataset, Data subjects can authorize a \textit{data controller} to access his or her personal data, with the possibility to transfer these to a \textit{data processor} in charge of processing these data. Another key actor is the \textit{Supervisory Authority} which is a controlling body. Each EU member state has its own National Supervisory Authority. In the context of scientific data sharing and reuse, individuals whose data records are collected by a researcher, are the data subjects and the researcher can be seen as a data controller. In case the researcher shares the dataset constituted by all the subjects’ data records and that another researcher reuses it, the second researcher can be seen as a data processor. The Supervisory Authority remains unchanged. 
Under GDPR data subjects have the right to obtain from the controller confirmation as to whether or not personal data concerning him or her are being processed~\cite{GDPR}. If that is the case, the controller also has to give the data subject access to her or his personal data, as well as some additional information such as the purpose of the processing and to whom their data has been transferred. The right to rectification states that the data subject has the right to have his or her personal data rectified by the controller, in case they are inaccurate. Finally, the data subject can also exercise a right to erasure where the controller is obliged to erase the subject's personal data. The fourth element of crucial importance, in the case of data sharing and reuse, is that of consent. Indeed, the GDPR requires that the data subjects freely give their informed consent to the collecting and processing of their data. 
Under this article, if personal data are shared and reused, it is vital that the purpose for which they are reused is compatible with the original purpose for which they were collected. 
As the primary data collector, the data provider remains responsible for the use of data and the main contact point for the data subjects. 

In the US, there is no specific federal data protection law, instead, it can be found in various legislations like Federal Trade Commission (FTC) Act~\cite{ftc}, Health Insurance Portability and Accountability Act (HIPAA)~\cite{hipaa} and California Consumer Privacy Act 2018 (CCPA)~\cite{ccpa}. The key privacy rights in these laws are the right to access the data or copies of these data, right to deletion of data, right to error rectifications, right to object to processing, right to data portability, right to withdraw consent, right to object to marketing and right to complain to relevant data authorities. Similarly, in UK, Data Protection Act 2018~\cite{uk} and in  Australia, Privacy Act 1988~\cite{au} promote the protection of data and privacy of individuals. In UAE, Dubai International Financial Centre (DIFC) has adopted the DIFC Law no. 1 of 2007~\cite{difc} and Abu Dhabi Global Markets (ADGM)~\cite{adgm} adopted the Data Protection regulations 2015 which are in line with EU GDPR regulations.


\section{LUCE Data Sharing Architecture}
\label{sec:architecture}
\begin{figure}[ht]
\centering
\includegraphics[width=0.85\textwidth]{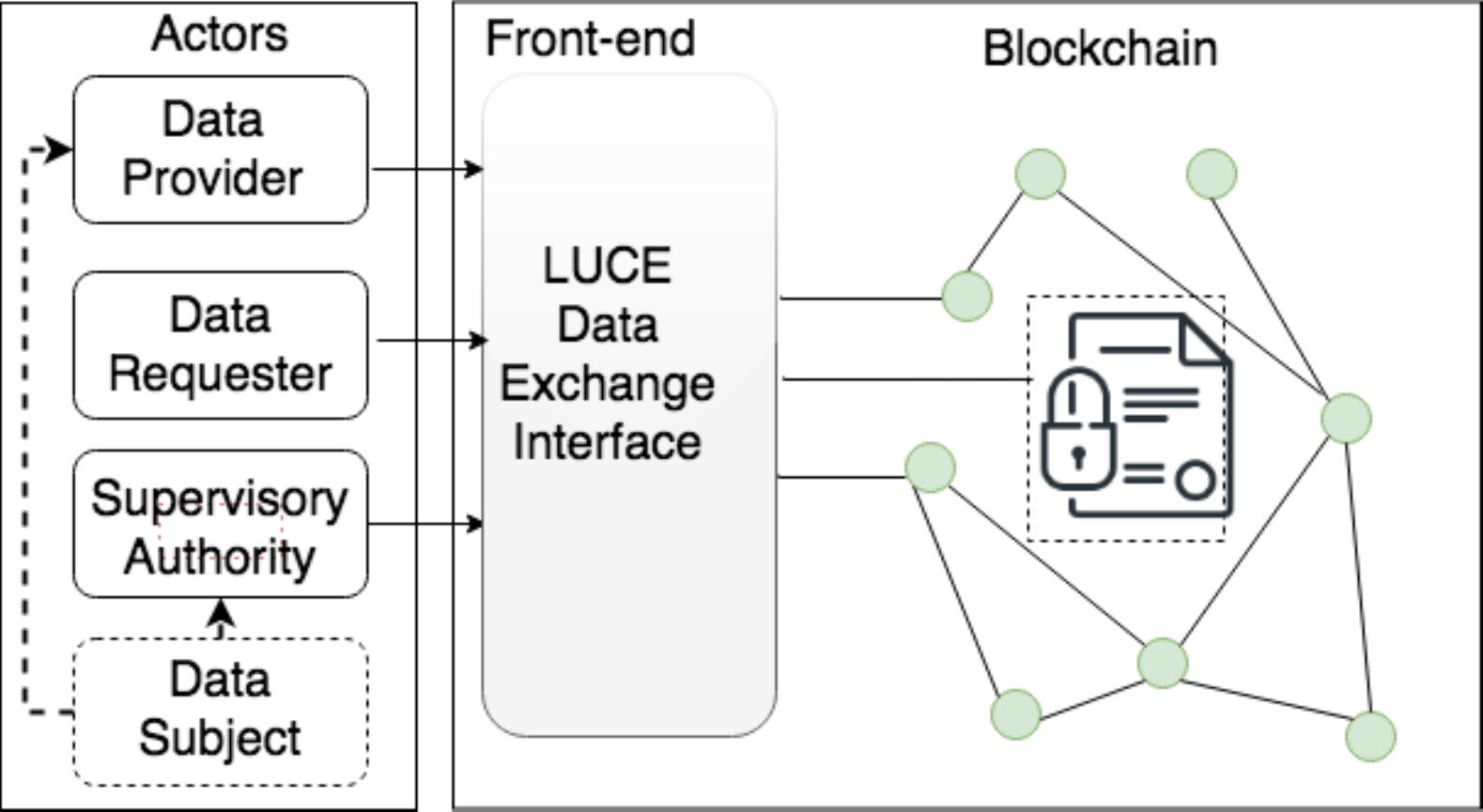}
\caption{LUCE actors and high view architecture.\label{fig:bigpic}}
\end{figure}

We define a model that i) manages licensing terms attached to a dataset, ii) dynamically manages user consent and reuse purpose of the dataset and, iii) enables compliance with the rights to access, rectification, and erasure of GDPR. Figure \ref{fig:bigpic} shows the main actors involved in data sharing and reuse, as well as how they interact. There are four main actors involved: the \textit{data provider}, e.g. a researcher willing to share a dataset, the \textit{data requester}, e.g. a researcher requesting to reuse a dataset, the \textit{supervisory authority}, e.g. a national public authority in charge of monitoring the adherence to data regulations and the \textit{data subjects}, e.g. any individual whose data are being collected, held, or processed. Data subjects trigger interactions via data providers or supervisory authorities, either by exercising their rights to access, erasure, modification, or by lodging a complaint via the supervisory authority. 
%
%

The interactions between data providers and data requesters are managed by smart contract agreements. These store information that is required to support data subjects to manage their data-sharing preferences and to exercise their GDPR rights. 
All the interactions in relation to data sharing are managed within the smart contract agreements and stored in an open blockchain network. The data themselves are not stored in the blockchain and token-based mechanisms are used to control the access to off-chain data. Each data provider, data requester, and supervisory authority is a node on a peer-to-peer blockchain network. LUCE checks that data requesters comply with the license of a dataset. For this purpose, data requesters are required to accept the licensing terms and to periodically renew their commitments within the smart contract. 
\begin{figure}[ht]
\centering
\includegraphics[width=\textwidth]{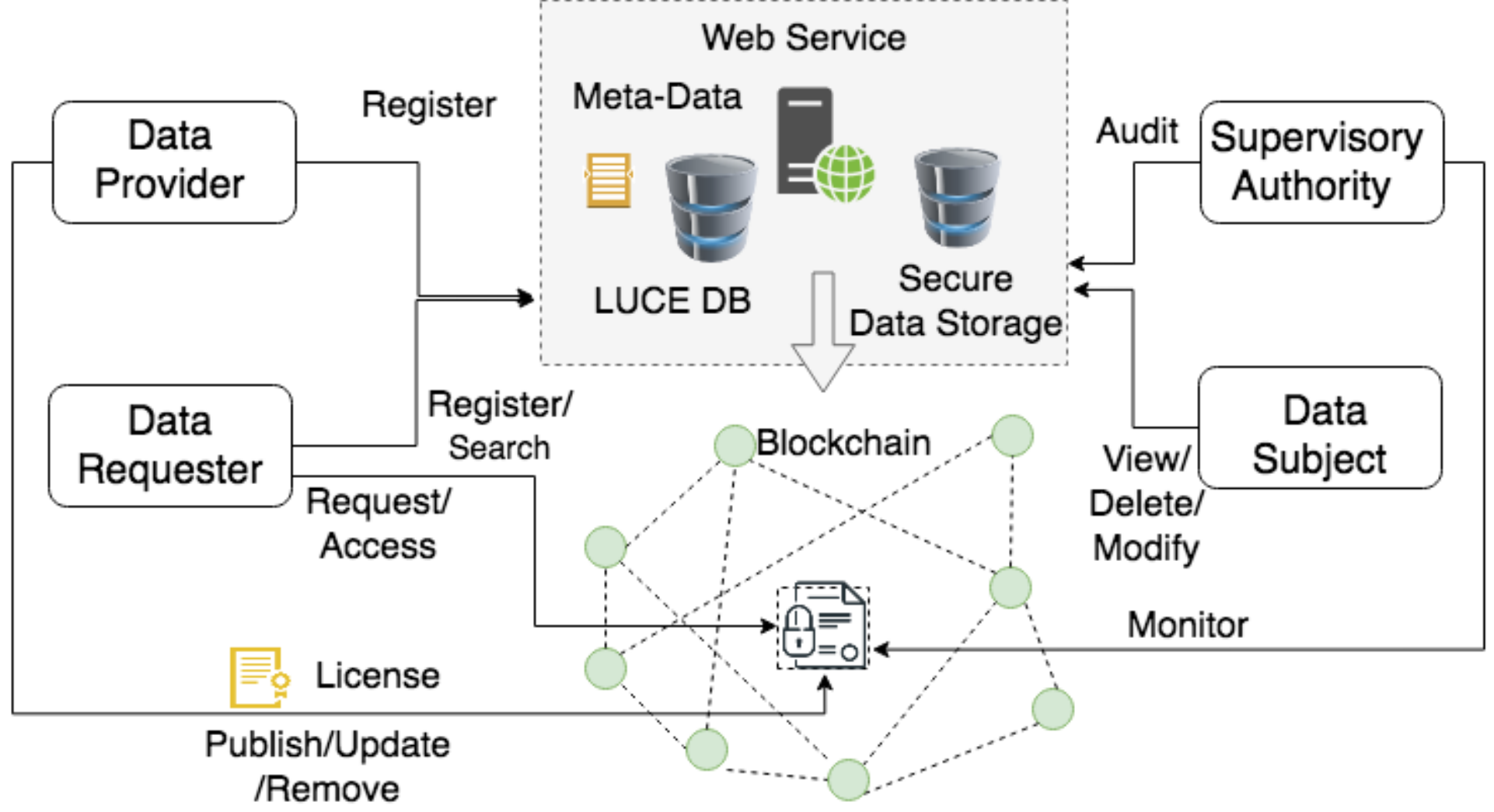}
\caption{LUCE architecture.\label{architecture}}
\end{figure}
LUCE focuses on three main data sharing steps: \textit{sharing data, reusing a data} and \textit{complying with the GDPR’s rights to access, rectification and erasure}. We make some assumptions regarding these steps:
\begin{enumerate}
\item Datasets are shared as a whole. 
\item Dataset integration is not taken into account, meaning that we currently do not cover cases in which several datasets are combined when reused. 
\item The license under which the dataset is shared is compatible with the consent of the data subjects. 
\item Personal data are de-identified and only the data provider can map data subjects to their records. The data subject and data provider can coincide when the individual is sharing their own data. 
\end{enumerate}
Assumptions 3 and 4 are only relevant in case the shared dataset contains data records of data subjects. All assumptions will be removed in future developments of LUCE. For example, in~\cite{jaiman2020consent} we already proposed an individual consent model for data subjects. In this model, data subjects can explicit their own consent over data sharing. In this way datasets are constructed based on the requirements of data requesters and the consent of data subjects.
Fig. \ref{architecture} shows an overview of the architecture of LUCE. In this diagram, the four primary actors – data provider, data requester, supervisory authority, and data subject interact with the components of the platform. In the following we explain the main components of Fig. \ref{architecture} and their interactions.

\subsection{Actors}

In LUCE, the owner of the data never changes. Data requesters subscribe to use a specific dataset and are encouraged to unsubscribe from it once its usage is no longer required. This minimizes the cognitive burden to the data requesters in that they only need to keep track of the licensing terms for the datasets currently in active use. It also gives data providers and the supervisory authority valuable chronological information on the time-frame and purpose for which a given dataset was used. If a dataset is to be re-used again at a later time or for a new purpose, access is requested again stating the new purpose. This keeps the purpose of use specific for any given access period. 
\par{}
A web interface facilitates interactions among users. The web interface offers different functionalities depending on the role of the user. The identity of the users is tracked via a user management system and each user identifies themselves to the system with their unique blockchain address \footnote{The following video link shows different LUCE actors interacting with the LUCE platform via the LUCE web interface \url{https://tinyurl.com/y22s9sq9}}. 
A \textit{Data Provider} can 1) \textit{Register} with the system 2) \textit{Publish} data 3) Update Data, and 4) \textit{Remove} data. Similarly, a \textit{Data Requester} is able to 1) \textit{Register} 2) \textit{Search} for data 3) \textit{Request} access to the data, and 4) Confirm license compliance periodically to maintain their access rights. The supervisory authority can access a mapping between registered users and blockchain addresses and access a list of all published datasets (past and present) and their corresponding smart contracts. Finally, the \textit{Data Subjects} are able to \textit{Register} as a data subject and \textit{View} which datasets have their information and for which purpose it is used.
\subsection{Smart Contract Agreements}
One smart contract is generated per published dataset. The smart contract: i) Keeps track of the access rights to the corresponding dataset. No data can be accessed without obtaining permission via the interaction with the smart contract.  Data requesters accept the licensing terms and enter a binding agreement with data providers. ii) Acts as proof of provenance. The hash of the dataset file is included in the contract so if competing claims of ownership arise, possession of the data at a specific point in time can be proven unambiguously by the data provider who published the contract to the blockchain. This provides a basic layer of protection against data theft and intellectual property infringements. iii) Creates an immutable history of who has had access to a given dataset, for how long, and for what purpose. This information is of interest to data subject who might exercise their right to information on how their data is being used. It is also used by the supervisory authority to audit whether actors comply with the licensing terms. 
\subsection{Maintaining the state in LUCE}
Several components were also introduced to maintain a local overview of the state of the platform. The \textit{secure Data Storage} component (see Fig.\ref{architecture}) is responsible for securely accessing the shared datasets. This component ensures that valid tokens, issued within LUCE are used to access data. 
A user registry contract keeps track of which blockchain addresses have been registered with the system. Only these addresses are allowed to interact with the dataset contracts. This way we can ensure that there exists a mapping between addresses and the real identity of platform users. All sensitive information pertinent to the users (data requester and/or data providers) is stored within \textit{LuceDB}. The smart contracts finally act as the gatekeepers to a dataset and ensure that access is only granted to data requesters who commit to adhere to the licensing terms. The smart contracts also contain a hash of the corresponding dataset, as well as metadata information, license type, permitted purposes, and a mapping of users with active access rights together with the  purpose for accessing the data. The smart contracts hold the ultimate reference of truth and can be publicly inspected if desired.
\textit{LuceDB} stores a synchronized version of the same truth which can be regarded as a cache. It allows for a seamless access to information via the interface without requiring to query all contracts on the blockchain about their current state every time that a user makes a request.

\subsection{Sharing a dataset}
\begin{figure}[ht] 
    \centerline{
    \includegraphics[width=0.75\textwidth]{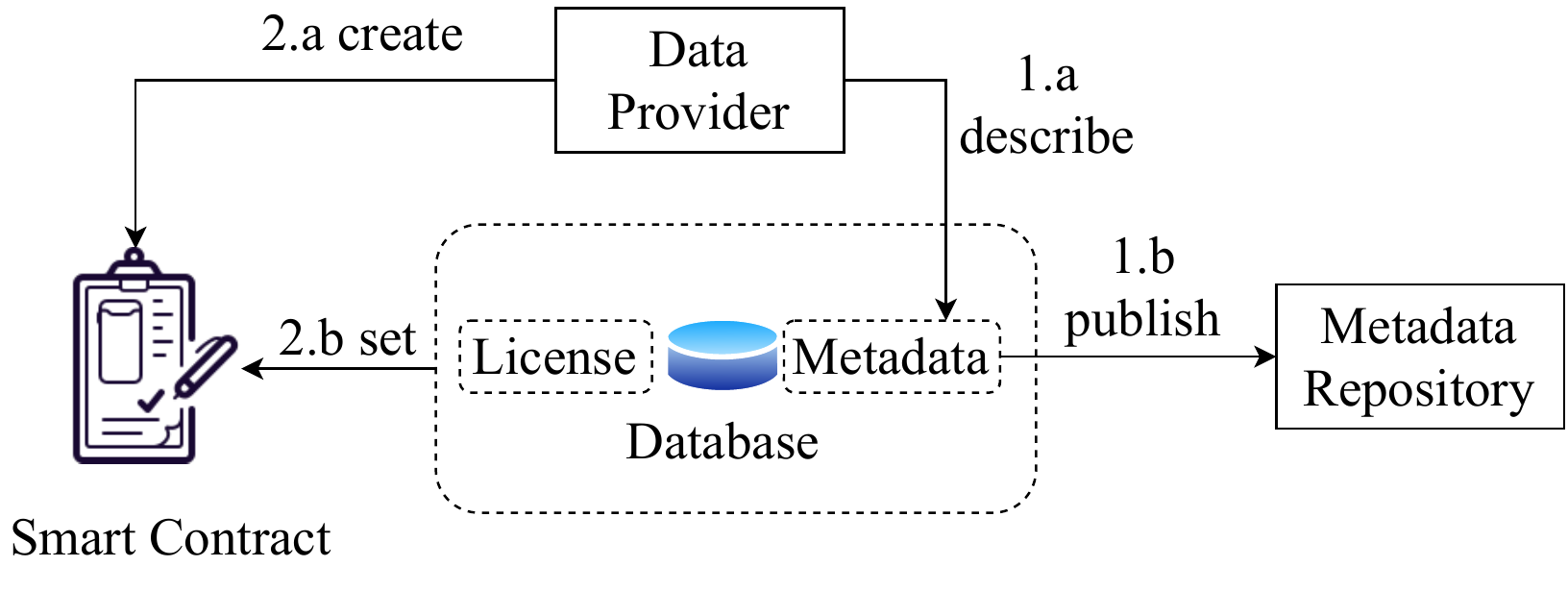}}
    \caption{Sharing a dataset}
    \label{Sharing}
\end{figure}

The first step in sharing data is a data provider sharing a dataset. Data are described and published as meta-data in the \textit{Metadata Repository}, an online shared directory where data requesters, can search for shared datasets. The shared meta-data includes the description of the dataset and the license attached to it. Uploading this information also creates the associated smart contract and the necessary information associated with the dataset. This sequence of interactions is presented in the diagram displayed in Fig \ref{Sharing}. 
\subsection{Reuse a dataset}
The second step in the process is to support data re-use. In this case a data requester wants to search, find and access a shared dataset. This happens in two steps. The first is to access the dataset and the second is the monitoring of compliance with the licensing terms of the dataset by the data requester.
\subsubsection{Accessing the dataset}
The first step in reuse a dataset is accessing it. Fig.~\ref{Reusing} illustrates the data access protocol to be used by data requesters. In particular, the steps below describe the process in detail:


\begin{figure}[!t] 
    \centerline{
   \includegraphics[width=0.75\textwidth]{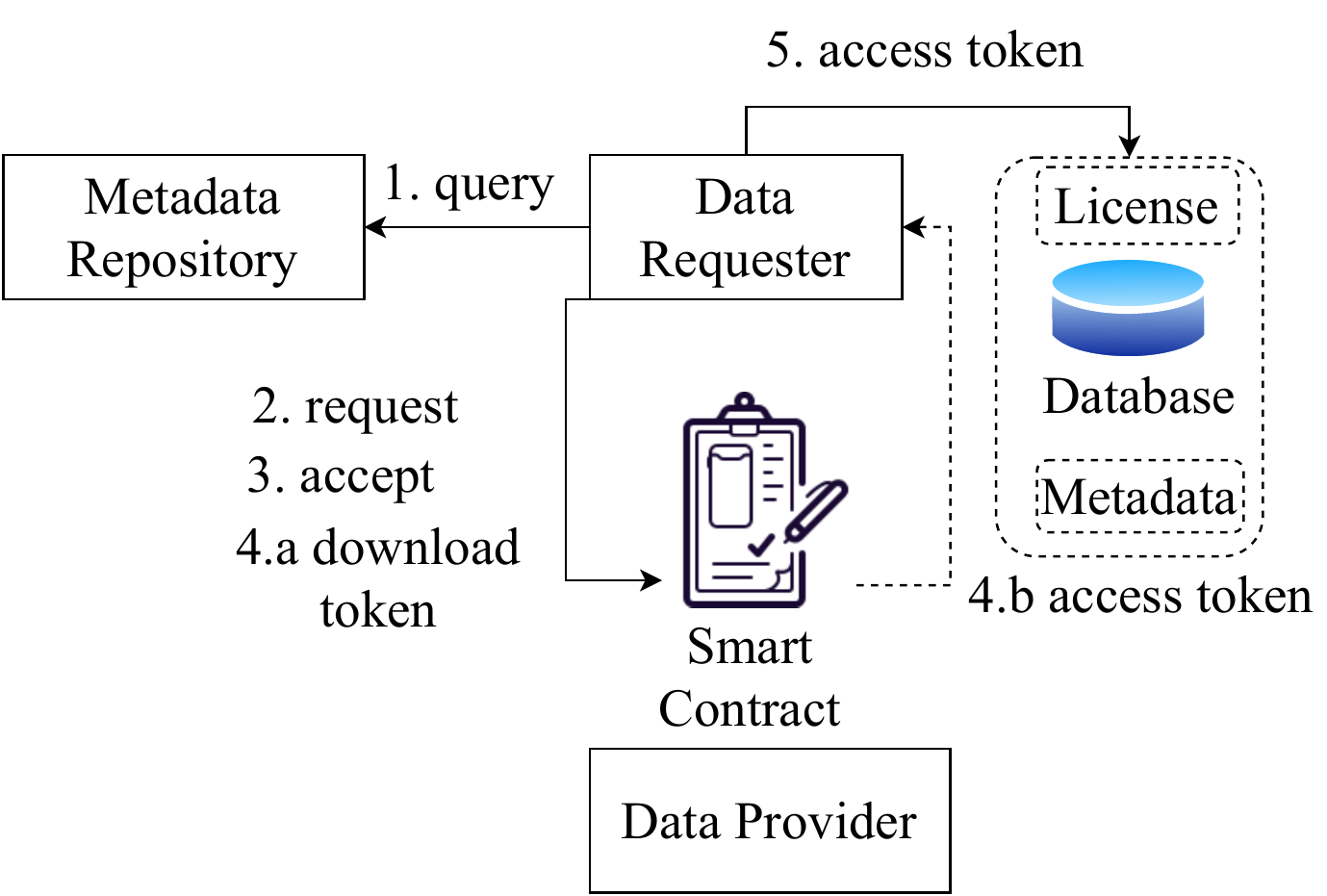}}
    \caption{Accessing a shared dataset}
    \label{Reusing}
\end{figure}
\begin{enumerate}
    \item \textbf{Query} Data requesters, who seek to reuse a dataset in the context of their own research, query the Metadata repository. For each dataset matching the query, the Metadata repository provides a list containing: the dataset identifier, its meta-data, the type of license attached to it and the smart-contract address. Based on this list, the data requester decides which database to reuse. As it has been discussed in~\cite{dubovitskaya2016multiagent}, more complex negotiation protocols on the data can be modelled at this stage.
    \item \textbf{Request} A request for accessing the dataset as well as the purpose of data reuse is sent to the smart contract by the data requester. In case the dataset contains data related to data subjects, the purpose indicated by the data requester is checked against the initial purpose these data were collected for. In case both purposes are compatible, the data provider grants access to the data requester. The compatibility of the data subject's consent and data requester's purpose is checked through a smart contract~\cite{jaiman2020consent}. 
    The specification of the purpose of reuse is also required in cases where the dataset does not contain data collected from data subjects.
    \item \textbf{Accept} The data requester is granted access to a dataset only after accepting the licensing terms. Thereafter, the data requester must provide information to the smart contract regarding the continued use of the dataset. 
    \item \textbf{Download token} After the data requester accepts the licensing terms and the check on the purpose of use, the smart contract provides the data requester with a verification token and a link to the repository where the dataset is stored.
\end{enumerate}

\subsubsection{Monitoring compliance with the dataset’s license}

\begin{figure}[!t] 
    \centerline{
    \includegraphics[width=0.75\textwidth]{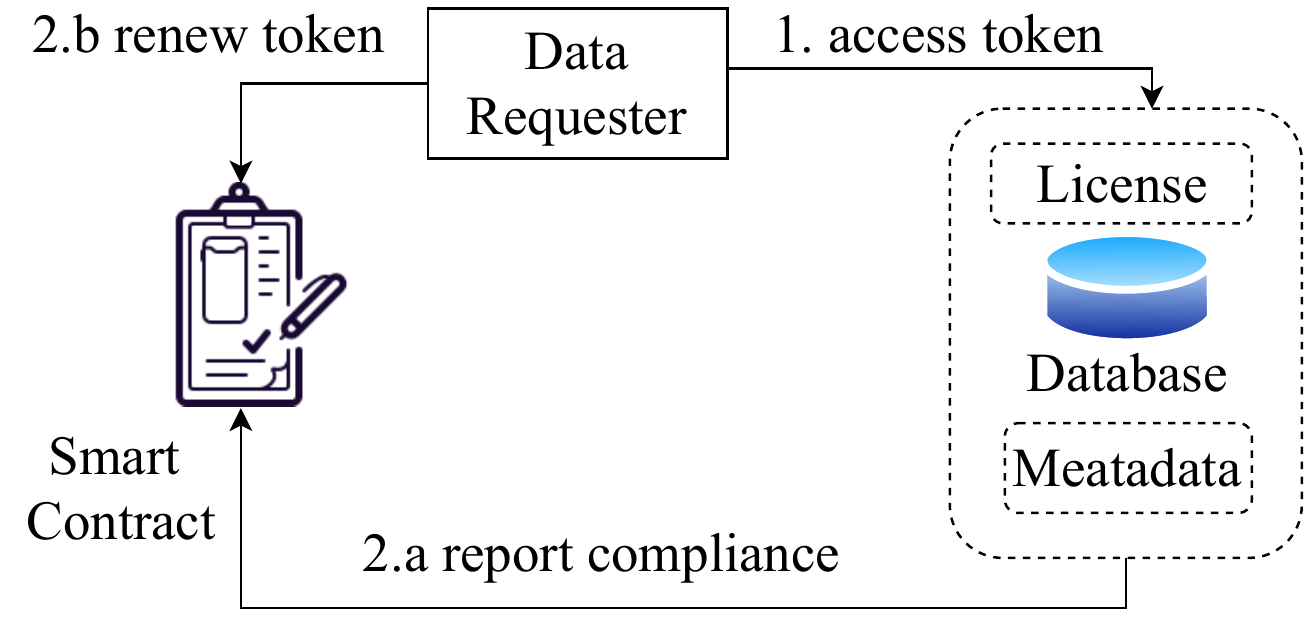}}
    \caption{Monitoring compliance with the licensing terms}
    \label{Compliance}
\end{figure}

The second step in reuse a dataset is the monitoring of compliance with its license by the data requester. This is illustrated in Fig. \ref{Compliance}. The following interaction is used to monitor compliance with the data:
\begin{enumerate}
    \item \textbf{Access token:} To access a given dataset, the data requester receives an access token. The access token is generated by the smart contract after the data requester accepts the licensing terms and describes the purpose of use for the data. The given token associates the token ID with the data requester. We extended the ERC-721 token~\cite{ERC721} to track the ownership of the data. This implies that data requesters can use tokens for three purposes: accessing the data, renewing access time to the data, and deleting their access to the data. This work is out of the scope of this paper and will be thoroughly presented in follow-up work.
    \item \textbf{Report compliance} Once a data requester obtains the access token, they can access the dataset. Two models of monitoring compliance are possible with LUCE: 1) creating an executable within a cloud-based model that monitors the modifications made on the dataset. This module continuously checks whether the actions performed by the data requester on the data comply with the licensing terms and records these actions in a file that constitutes a log of events, or, 2) Supporting a publish-subscribe model where users show/declare that they comply with the rules by periodically sending transactions to the contract. In this way, we are not fully checking license agreements, but we have continuous checkpoints and the public commitment of data requesters towards licensing terms. This model allows data requesters to download the dataset and requests frequent renewal of the token for continued use. The first option implies that data reuse and analysis is possible only on provided software. The second model implies a more flexible data reuse as long as the token is renewed. LUCE requests the renewal of the token for continued use, independently from which compliance model is used. 
    \item \textbf{Renew token} The access token is used by the data requester to obtain access to the dataset. The token is only valid during a period of duration T. When that period ends, the data requesters renew their access token by requesting the renewal to the smart contract. The smart contract verifies whether the data requester has complied with the licensing terms during the last period. If so, the contract renews the access token to the data requester. However, in case the licensing terms were not complied with, the access token is not renewed, meaning that the data requester cannot access the dataset anymore. 
\end{enumerate}
Independently from the chosen monitoring model, a data provider can check whether data requesters have complied with the license terms attached to a shared dataset by accessing the events in the smart contract associated with the dataset.

\begin{figure}[ht] 
    \centerline{
    \includegraphics[width=0.75\textwidth]{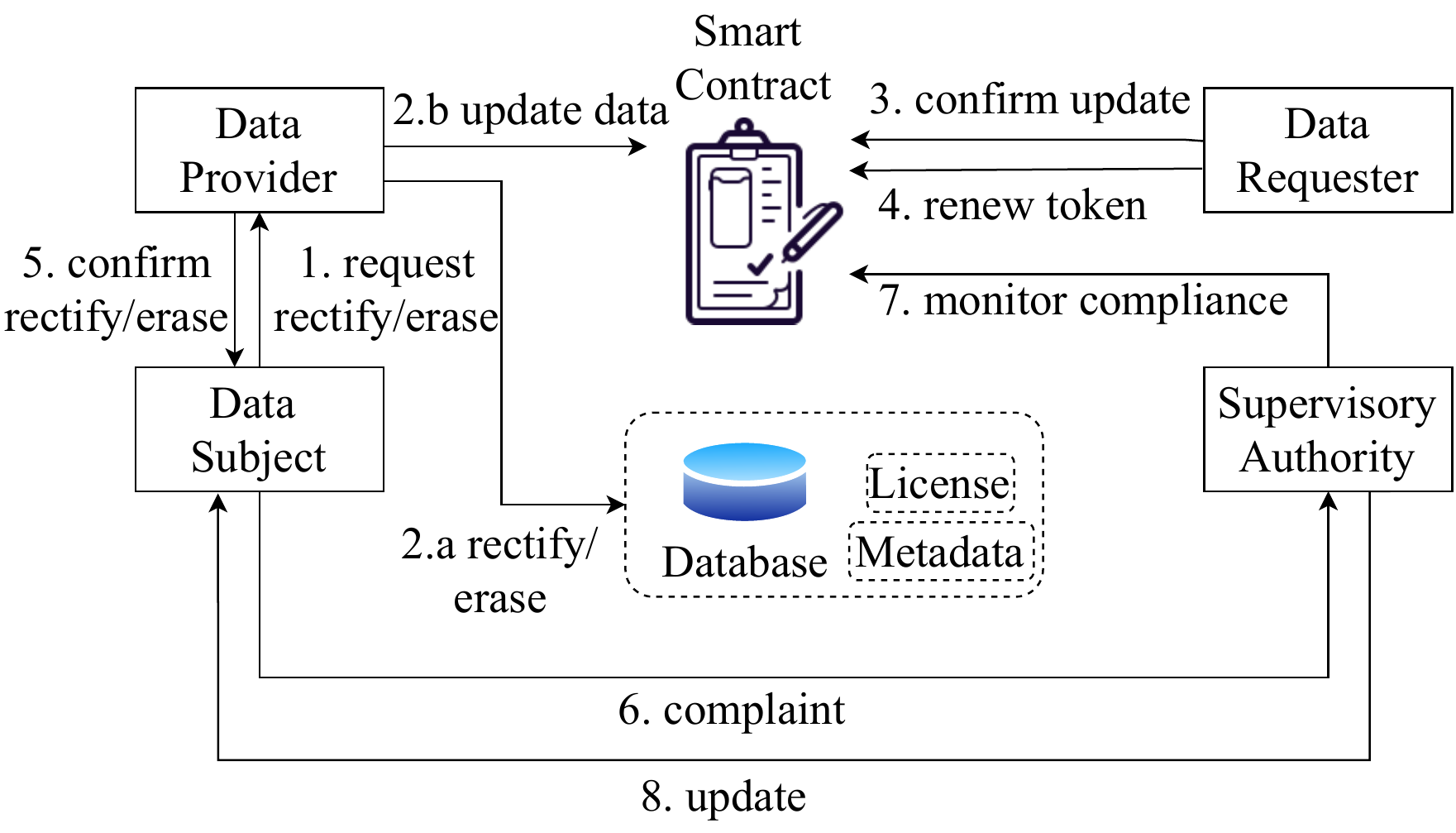}}
    \caption{Complying with GDPR's rights to access, rectification and erasure}
    \label{GDPRsimple}
\end{figure}

\subsection{GDPR compliance}
The third data-sharing step covers the cases in which a data subject exercises their right to access, erasure, or rectification. Fig. \ref{GDPRsimple} shows how we model the interactions with regards to the rights to rectification and erasure. In particular, the interaction steps can be described as follows: 
\begin{enumerate}
\item \textbf{Request rectify/erase:} 
In concordance with the GDPR, the data subject has the right to request information from the data provider as to how their data is used, by whom, and request changes or deletion of their records. Data providers are the only ones that hold a mapping between a data subject and the dataset in which their records are included. At any time a data requester can query the smart contract to identify who is reusing the data and for what purpose and provide this information to data subjects. In case the data subject requests the data provider to either rectify or erase their data from the dataset, it must be done not only in the dataset of the data provider but also in the copies of the dataset that data requesters are using. 
\item \textbf{Data updates:} consists of two actions
\begin{enumerate}
\item \textbf{Rectify/erase:} A data provider must anonymize the data before sharing them for further reuse. The mapping of data subjects to anonymized IDs is carefully stored by data providers (this information is not shared with the LUCE platform). After the request, a data provider uses the anonymized ID of the data subject to modify or erase the data. New data requesters will only access the updated records. 
\item \textbf{Update data:} The smart contract contains all the identifiers of data requesters who are re-using the dataset. Using the anonymized identifier of the data subject, the erasure or modification of specific records is required for all the data requested who are re-using the data. For this, the smart contract generates an event for all the data requesters who are currently using the data.
\end{enumerate}
\item \textbf{Confirm update:} All the data requesters who are using a non-updated copy of the dataset must erase or modify the data records as requested. The change is confirmed via a confirm update message sent to the smart contract.
\item \textbf{Renew token:} The data providers must renew their token after an update. This operation ensures that the data requesters who do not comply with the request are revoked access to the data.
\item \textbf{Confirm rectify/erase:} The data provider queries the smart contract to collect the information about the data requesters who are reusing the data and the state of the update. All the information is sent to the data subject. The confirmation of a rectify/erase has a delay which is dependent on the frequency of token renewal. From a data subject request of a rectify/erase the data subject waits at maximum the time of a token renewal. Such parameter needs to be carefully decided to ensure a good equilibrium between the expense that occurred by performing frequent transactions in the system and the accuracy of information to be provided to data subjects and supervisory authorities. In our current implementation, the frequency of token renewal is set to 2 weeks.
\item \textbf{Complaint:} Should the request of a data subject not have a satisfactory outcome, a complaint to the supervisory authority can be made. 
\item \textbf{Monitor compliance:} LUCE enables the supervisory authority to check for GDPR compliance. The supervisory authority can monitor all the transactions of the blockchain platform. With the information provided by the data subject, it can verify the actions of data requesters in the corresponding smart contract. They also might collect evidence documents as well as check the behavior of a data requester towards multiple other datasets or data subjects.
\item \textbf{Update:} After monitoring compliance, the supervisory authority notifies the data subject about the outcome.
\end{enumerate}
\section{Implementation}
\label{sec:implementation}
In this section, we will provide the details regarding the implementation and the experimental setup of LUCE and a proof of concept web portal to access and share data on the LUCE platform. Moreover, we will discuss the LuceVM -- a virtual machine implementation of LUCE and LuceDocker -- a dockerized version of LUCE to illustrate the usage of the LUCE platform. 
\subsection{Implementation and Experimental setup}

We implement LUCE on top of Ethereum~\cite{ethereum} blockchain, an open-source and public blockchain-based distributed computing platform for building decentralized applications. Choosing the Ethereum platform implies that LUCE’s blockchain can be defined as a public platform, and anyone can inspect transactions as well as add or verify new ones. 

While it is technically possible for users to interface directly with Ethereum, LUCE provides a user-friendly web-based solution. For the development of the web platform, we use the python Django framework~\cite{django}, and integrate it to Ethereum via web3~\cite{web3js} and py-solc. The platform includes a web interface as well as an application programming interface that allows for programmatic access to the system. Thus, LUCE users can interact with the platform via the LUCE Data Exchange web interface or programmatically via a set of functionally equivalent APIs. 
The back-end encapsulates a Web Server (WSGI) and the application code which together provide the business logic. The database (LuceDB) handles user account management and acts as a cache for metadata to feed the web interface.

To run our experiments, we run LuceVM virtual machine on a 64 bit Ubuntu 16.04 LTS (Xenial Xerus) Linux operating system. The virtual machine is equipped with 4096 MB RAM. The virtual machine is managed by Vagrant~\cite{vagrant} to further abstract away the layer of VM configuration for the end-users. Moreover, the VM is preconfigured with the blockchain development environment, libraries, and connected servers. Our LUCE platform implementation is available as open-source~\footnote{https://github.com/vjaiman/LUCE}. 

For fast deployment of LUCE, we use Docker~\cite{docker} to bundle the LUCE software. The dockerized image of LUCE is deployed on a server hosted by the Institute of data science, Maastricht University. 
With this web-hosted version, a user can register with LUCE and share the dataset on the platform.
\subsubsection{LUCE Smart Contract}
\begin{figure*}[!t]
\centering
\includegraphics[width=\textwidth]{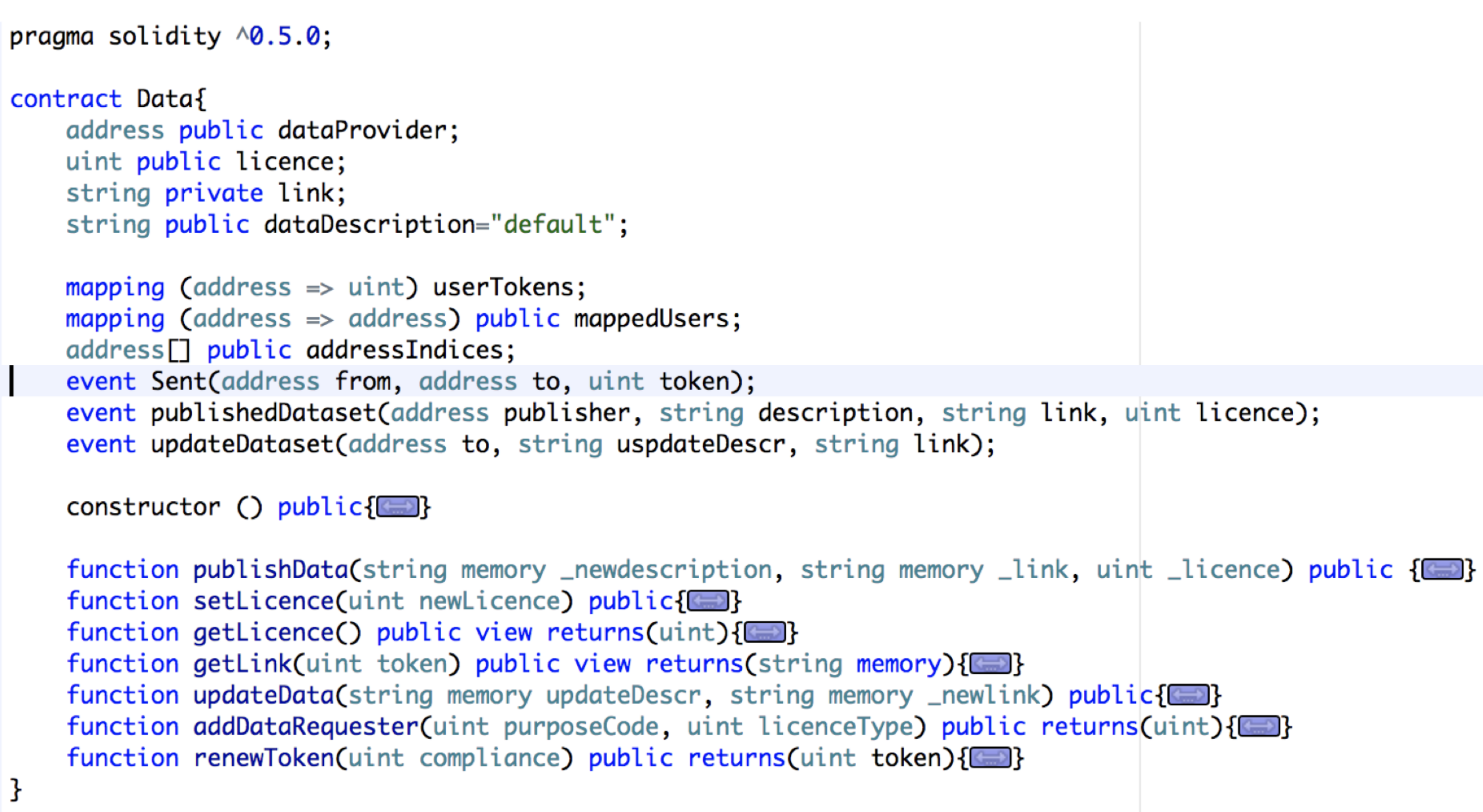}
\caption{Smart contract model.\label{codesnippet}}
\end{figure*}
As previously described, smart contracts~\cite{smartContract} capture the interactions between data providers and requesters. Each data provider has a smart contract that is used to publish the dataset on the platform with preset conditions. Fig.~\ref{codesnippet} shows the code snippet of our smart contract used in the LUCE platform. 
When publishing a dataset, a data provider sets within the smart contract the data sharing conditions on the LUCE platform (\textit{publishData}). In doing so, a description and a persistent URL to the data is provided by the data provider.
Data requesters can use a shared dataset by calling the \textit{addDataRequester} function. To call this function, they would have first to read the license terms with the \textit{getLicence} function.
For continued access to data, tokens must be renewed via the \textit{renewToken} function. If data requesters comply with the license, then the token will always be renewed, otherwise the token is revoked and so is the access to the data.
The interactions with the smart contract generate transactions that are added to the blockchain. It takes some time to add a transaction in the Ethereum blockchain that typically ranges between $\sim$10-20 seconds. We test LUCE by simulating interactions within the platform. All our simulated experiments were repeated 4 times to draw an overall average. We set up our experiments on private, and a public blockchain testnet.
\subsubsection{Private Testnet}
For testing purposes, Ethereum enables users to set up private blockchains that are separated by the Ethereum mainnet. This solution is used to test the blockchain environment before exposing it to the Ethereum mainnet where real costs occur (in the form of ether, the Ethereum currency). We make use of geth~\cite{geth} to connect our node to the testnet. We conduct the experiments on a private blockchain testnet by creating 6000 accounts on it. Each account has to be unlocked before initiating any transaction and having an associated address to it by which it can be identified. Moreover, each account can hold  \textit{ether} to bear the transaction cost. In a private blockchain, ether is received for \textit{free}. 
\subsubsection{Public Testnet}
We also tested LUCE on a public testnet. Public testnets are very similar to the Ethereum mainnet network in terms of operation (i.e. transaction times are of a dynamic length and very similar to the mainnet. This is not the case with a private testnet.). The difference is that while Ether must be acquired to operate on the main Ethereum network, in public testnets Ether remains free. 
We test the LUCE on the public Rinkeby testnet~\cite{rinkeby}. We make use of geth~\cite{geth} to connect our node to the Rinkeby testnet. We conduct our experiments by creating 2000 accounts on the public network. The coinbase account can be funded through \textit{https://faucet.rinkeby.io} for 3 ether by publishing the coinbase address on Twitter or Facebook. All the pending/completed transactions are available at Rinkeby Etherscan explorer~\cite{rinkebyetherscan}. The coinbase address is ``0x9f913ef90c695ae1529e6cf5f1a5d407fe1a4178" and all the completed/pending transactions can be viewed by coinbase address on Rinkeby Etherscan explorer~\cite{rinkebyetherscan}. 
\section{Evaluation}\label{sec:evaluation}
We evaluate the effectiveness of our solution in terms of scalability and cost, 
We compare the cost with a \textit{baseline} smart contract performing minimal operations i.e. to set the input at a given address. The code of the baseline smart contract is publicly available~\footnote{https://github.com/vjaiman/LUCE}. Moreover, we compare it with consent-based LUCE~\cite{jaiman2020consent}, a more complex model where a data requester gets access to the data only if the purpose of use of a data requester and the consent use of a data provider match. The consent-based model adds to the licensing terms with more detailed information on the purposes under which data reuse is allowed (This model was presented in ~\cite{jaiman2020consent}). Our evaluation aims to answer the following questions:
\begin{enumerate}
    \item How well does the data sharing model of LUCE scale with an increasing number of data requesters requesting the same dataset?
    \item How the data sharing model scales with an increasing number of data requesters requesting different datasets?
    \item How cost-effective are the operations in the LUCE platform?
\end{enumerate}
The first question analyses how well LUCE scales with an increasing traffic on a single smart contract and the respective dataset. The second question looks at how well LUCE scales with increasing traffic in general and the third questions analyses the transaction costs introduced by user interactions.

\begin{figure}[!t]
\centering
\includegraphics[width=0.85\textwidth]{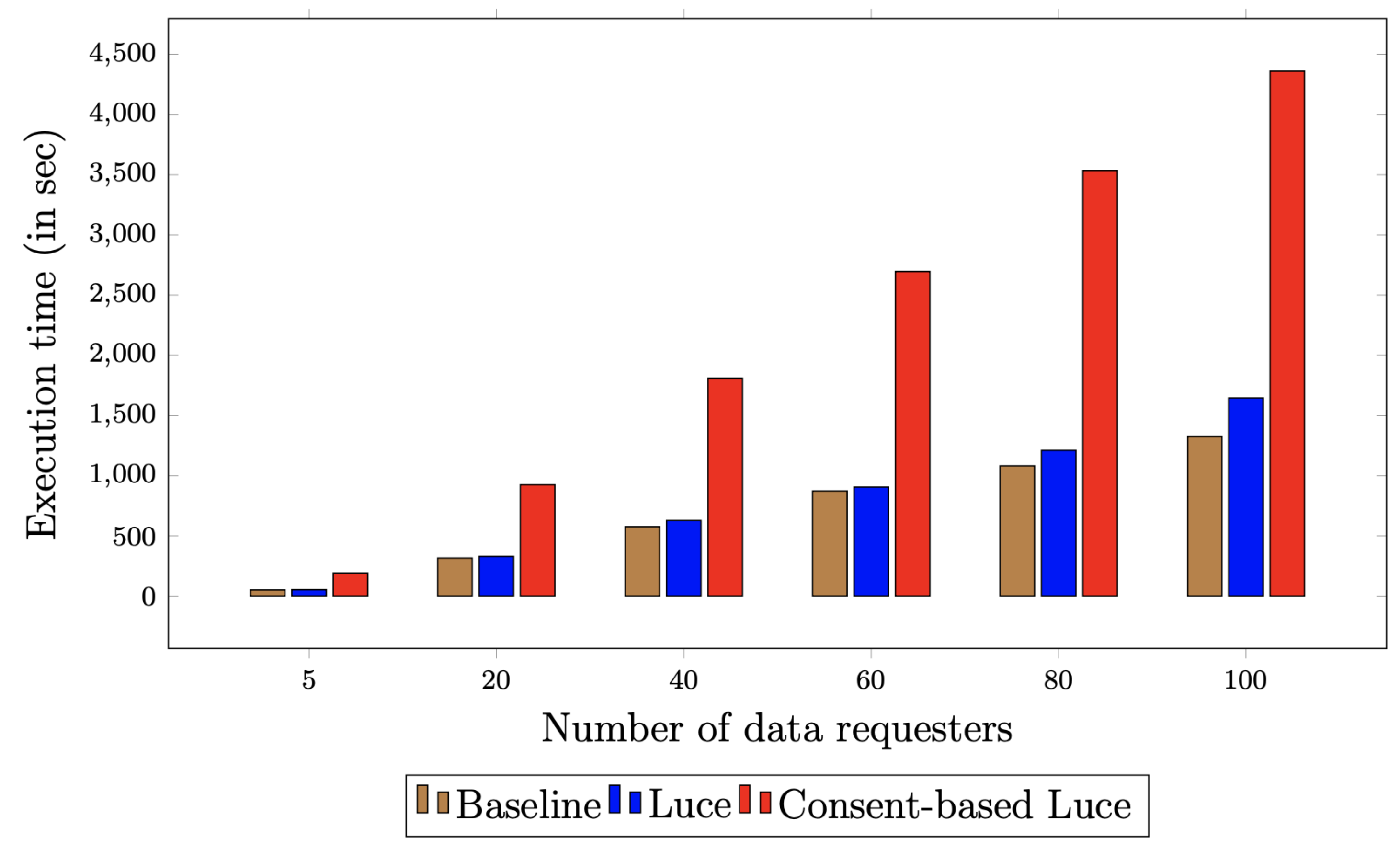}
\caption{Mining latency of data requesters in LUCE on private testnet.\label{fig:mininglatencyprivatetestnet}}
\end{figure}

Fig.~\ref{fig:mininglatencyprivatetestnet} shows the execution time in LUCE with an increasing number of data requesters (5-100) who request access to the same dataset.  We calculate the execution time of mined transactions. In general, mining a transaction takes between 10-20 seconds and can be longer depending on the network congestion and the gas price. In this experiment, after submitting a transaction, we wait for the transaction receipt from the miners. Afterwards, we calculate the total execution time taken by LUCE and Ethereum private testnet. We observe that, as the number of data requesters increases, the total execution time is almost the same in LUCE compared to the baseline. We can observe a little difference once data requesters increase to 80-100 which might be explained by the list scrolling operations of the smart contract to identify that transactions initiated by data requesters went through the required interaction protocol. Overall the results show that the smart contract model of LUCE introduces low computational cost, compared to the transaction cost introduced by the architecture of the Ethereum blockchain. Compared to the Consent-based LUCE~\cite{jaiman2020consent}, the second model takes a longer time to execute as the smart contract performs several checks to determine that the purpose of use matches a given consent. This second model suggests that consent-based access to the data may further increase the latency of transactions, especially when multiple concurrent data access queries are made on the same dataset.
\begin{figure}[!t]
\centering
\includegraphics[width=0.85\textwidth]{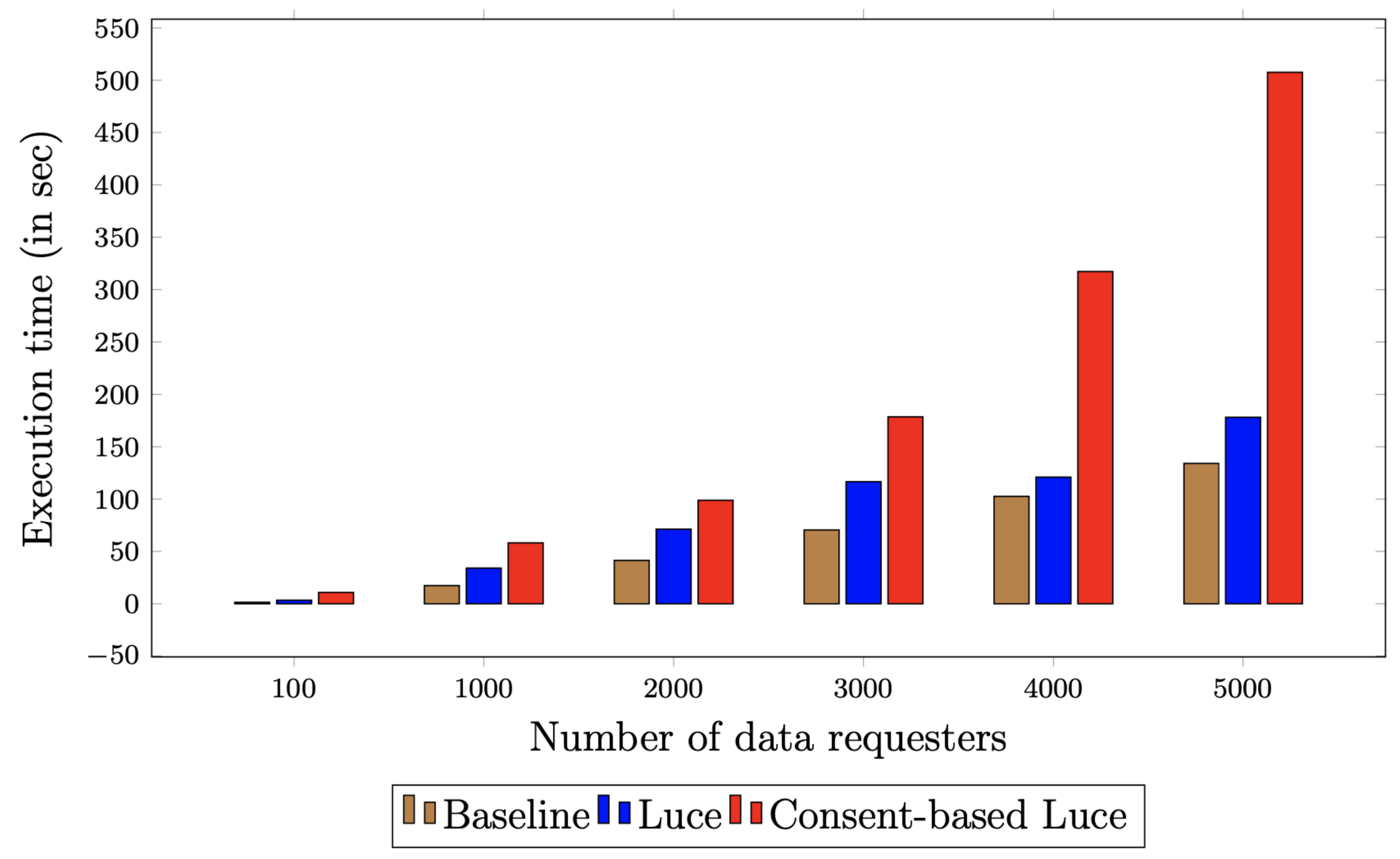}
\caption{Latency of data requesters on the private testnet.\label{fig:latencyprivatetestnet}}
\end{figure}
Fig.~\ref{fig:latencyprivatetestnet} shows the execution time with an increasing number of data requesters accessing data in the LUCE platform. In this experiment, we submit transactions from an increasing number of data requesters (100-5000). Contrary to the previous experiment in Fig.~\ref{fig:mininglatencyprivatetestnet}, we do not wait for the transactions to be mined by Ethereum and calculate the execution time for submitting the transactions by the LUCE platform. The objective was to see the actual execution cost of functions defined in the smart contract. In this experiment, LUCE smart contract takes 178.09s whereas \textsf{baseline} smart contract takes 134.08s. We can observe that the smart contracts of LUCE impose low operational cost for data sharing compared to the baseline smart contract. In fact, consent-based LUCE takes 507.55s due to the submission of consent before getting access to the data.
\begin{figure}[!t]
\centering
\includegraphics[width=0.85\textwidth]{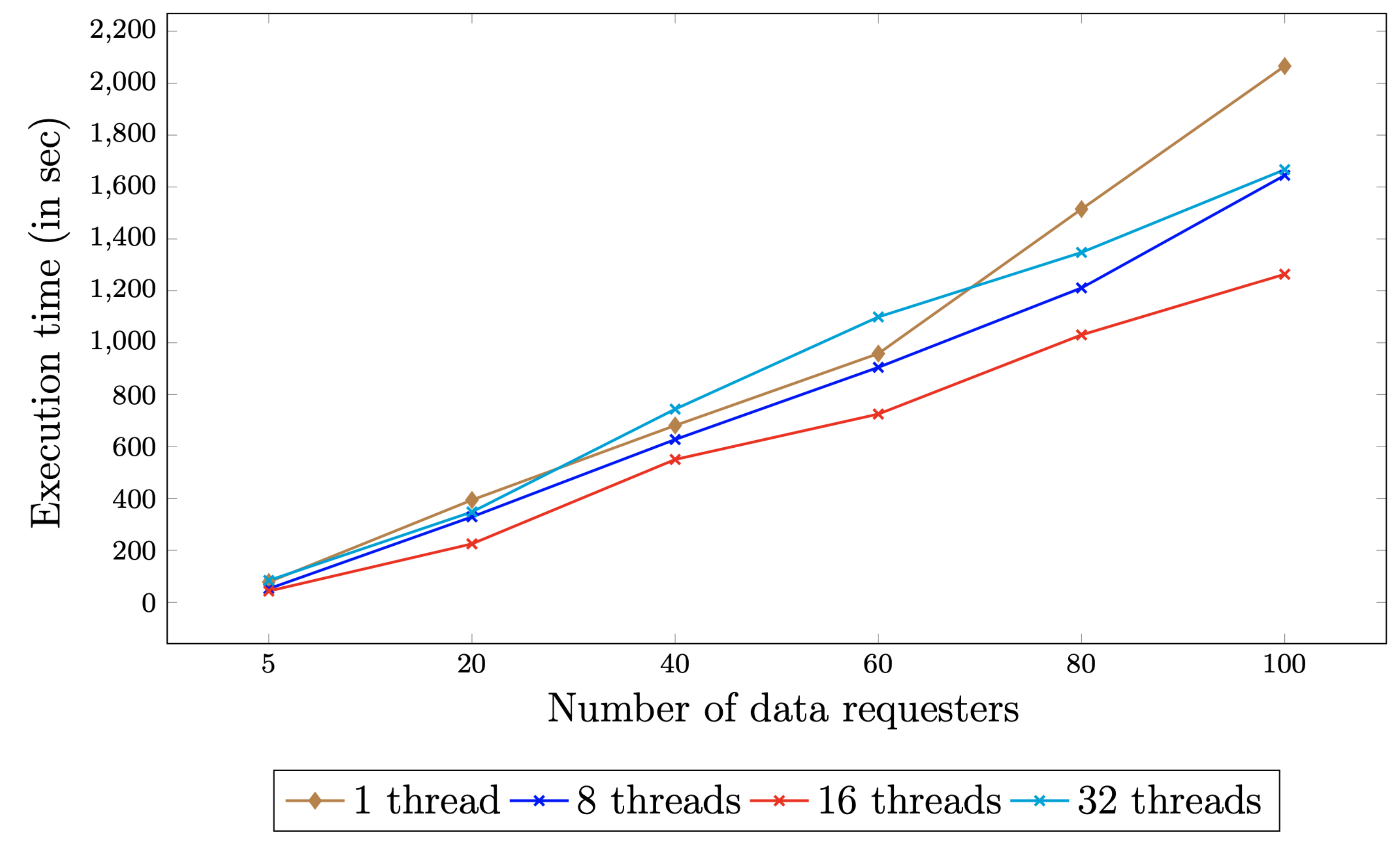}
\caption{Impact of varied mining threads on the private testnet.\label{fig:miningthreads}}
\end{figure}
In Fig.~\ref{fig:miningthreads}, we vary the mining threads to see the impact of miners on the incoming data requests. For 100 data requesters, with 16 threads, it takes an optimum time of 1264s whereas with 1 thread it takes a maximum time of 2066s. We observe that as the number of data requesters increases the mining time in each thread group also increases. Moreover, mining time drops from 1 to 16, and afterwards due to thrashing and network congestion, mining time increases again for 32 threads. We conclude that under these settings, LUCE works efficiently with more miners when the number of data requesters increases. 
\begin{figure}[!t]
\centering
\includegraphics[width=0.85\textwidth]{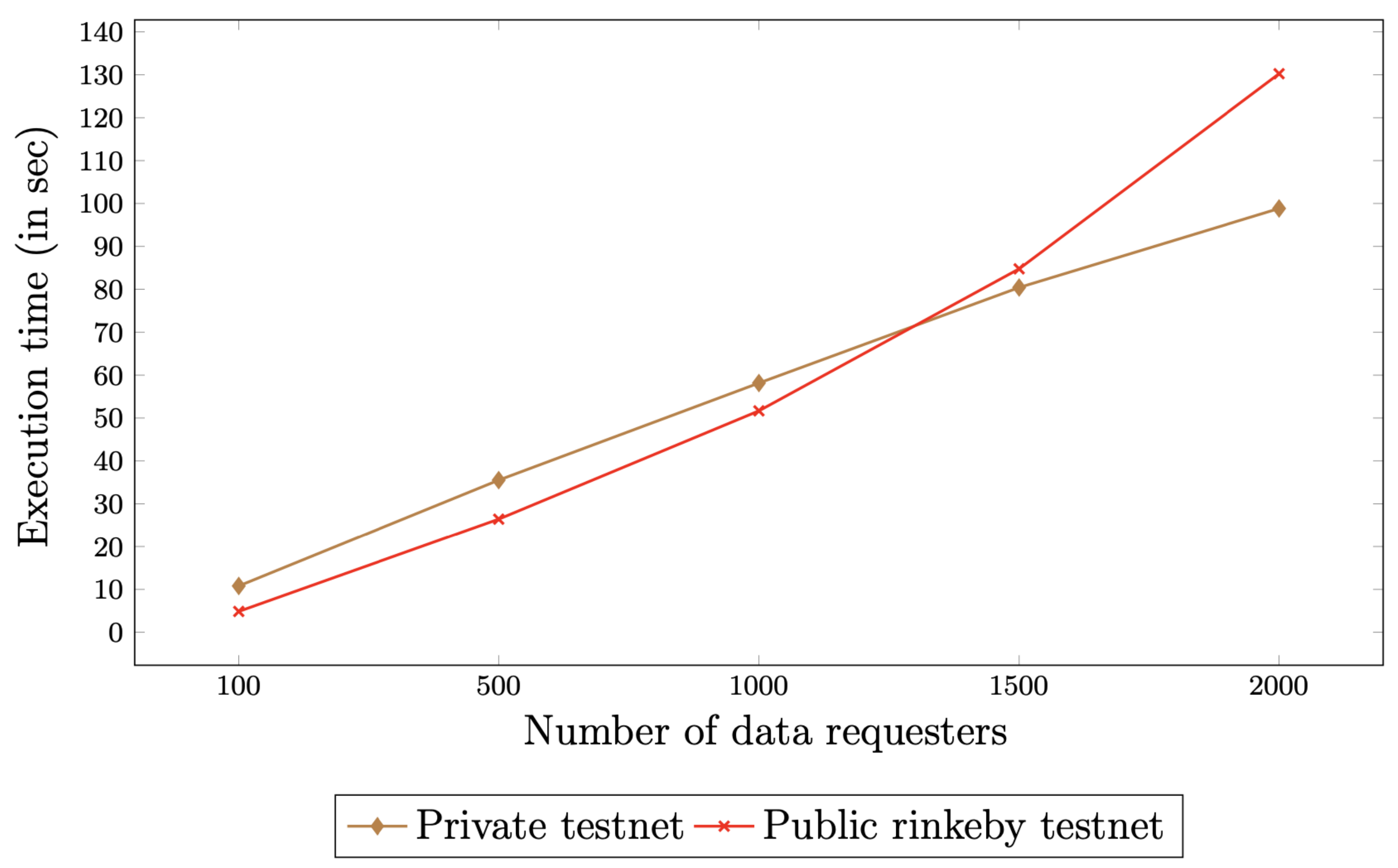}
\caption{Latency of data requesters in LUCE on private testnet and public rinkeby testnet. \label{latencyrinkeby}}
\end{figure}
For public testnet experiments, we submit transactions on Rinkeby testnet by increasing the data requesters from 100 to 2000 for the consent-based LUCE. The objective is to observe the execution time difference for LUCE compared to the private testnet. In Fig.~\ref{latencyrinkeby} we observe that consent-based LUCE takes 51.67s for 1000 data requesters compared to the 58.15s for private testnet. Similarly, the public testnet takes 130.25s for executing the 2000 data requesters compared to 98.84s for the private testnet. We observe that, when the number of transactions increases, the transactions create a congestion of the public testnet resulting in a longer time to execute compared to the private testnet. In the following section, we will analyze the corresponding cost to evaluate the LUCE platform.
\subsection{Cost analysis and feasibility}
In Ethereum, every transaction in the smart contract consumes gas (a measure of the computational effort required to perform an operation). Gas consumption varies based on the complexity of the functions defined in the smart contract. We perform a cost analysis of the functions defined in the LUCE smart contract. Several parameters are considered:
\begin{enumerate}
\item The total amount of gas spent during LUCE contract deployment.
\item The amount of gas consumed when publishing the dataset and setting the license.
\item The amount of gas consumed when updating the data or renewing the token.
\item The amount of gas spent while getting the license and the token.
\end{enumerate}
 Table~\ref{table:costanalysis} shows the total gas consumption to execute the contract is 1339598. We considered the average gas price of 32 Gwei according to the current date\footnote{16/07/2021} of ETH gas station~\cite{gasstation}. Therefore, the relevant cost of contract deployment is 0.0428671 ETH with a corresponding price of \$79.28 (1 ETH == \$1849.44). It is noted that the transaction cost includes the default cost of 21000 gas which is occurred by Ethereum for performing a transaction on it and therefore the actual function cost is mentioned as \textit{Execution cost} in the table~\ref{table:costanalysis}.  
\begin{table*}
\caption{Base cost for the core functions of LUCE.}
\label{table:costanalysis}
\setlength{\tabcolsep}{6pt}
\def\arraystretch{1.5}
\begin{tabular}{p{110pt}p{85pt}p{85pt}p{85pt}p{85pt}}
\hline
 Actions & Transaction cost (in gas) & Execution cost (in gas) &  Ether cost (in ETH) & Cost* (in \$) \\
\hline
Contract\\Deployment & 1339598 & 964030 & 0.0428671 & \$ 79.28 \\
\hline
publishData & 79652 & 56460 & 0.002548 & \$ 4.71 \\
\hline
setLicense & 24201 & 2737 & 0.0007744 & \$ 1.43 \\
\hline
addDataRequester & 105842 & 84186 & 0.003386 & \$ 6.26 \\
\hline
updateData & 47756 & 24884 & 0.001528 & \$ 2.82\\
\hline
renewToken & 16149 & 9685 & 0.0005268 & \$ 0.97\\
\hline
getLink & 24780  & 3316 & 0.000793 & \$ 1.46\\
\hline
getLicense & 22384 & 1112 & 0.000716 & \$ 1.32\\
\hline
\multicolumn{5}{p{430pt}}{*= Ether conversion with present date price (Average = 32 Gwei \& 1 ETH = \$1849.44)}
\end{tabular}
\end{table*}
We calculated the cost of data sharing by the accumulative cost of contract deployment, the cost for publishing the dataset, and the cost of setting the license. Some additional costs are involved when data is updated or the token is renewed. We observe that publishing a dataset and adding a data requester involves more gas consumption compared to the other operations. This is due to the involved parameters, such as setting the description of the dataset or setting up the link to the dataset location. Similarly, adding a data requester requires mapping a new user against the existing ones, therefore, such operation requires more computational efforts and gas consumption.

\section{Related Work}
\label{sec:related}
The technological advancement of the last few decades has enabled collaborations and markets for sharing data. In the current practices,  personal data from different data sources are collected and combined (often without explicit user consent) with the purpose of aiding data-driven decisions. This way of sharing and re-using personal data has brought important legal and social implications. In response, several decentralized solutions to data sharing have been proposed (for example~\cite{rajput2021blockchain, chi2020secure, shrestha2020blockchain, patel2019framework, jaimanincentive}). The overall aim of these solution is similar to LUCE, decentralize the control over data to bring more control to the users and support transparency over data exchange.
In particular, in \cite{rajput2021blockchain} and in~\cite{patel2019framework} the authors define  permissioned blockchain solutions that enables respectively sharing of personal health records in emergency situations and cross-domain image sharing. In~\cite{rajput2021blockchain}, the authors focus on securing and preserving transactions from being tampered. This is done by means of smart contracts in which patients can enable access for specific doctors. While in~\cite{patel2019framework} the authors have a theoretical model defining how concepts such as the definitions of the study, the source and access mechanisms work towards data sharing.  In this work we seek not only to enable data-sharing but also monitor their use.
Moreover, while these works, and many others that could have been mentioned here (such as~\cite{azaria2016medrec, shen2019medchain, liang2017integrating}), focus on solving some aspects of data sharing and specializing the solution for the given scenarios, they do not fully address data sharing as a more generic process, requiring data to be shared but also updated maintained and eventually deleted.  
Some works have specifically focused on data provenance~\cite{Ramachandran2017, Liang2017, Liang2017} which describes the history of data, such as where they originate from, their owner, the changes made to the data, as well as who made them~\cite{Ramachandran2017,Liang2017}. DataProv~\cite{Ramachandran2017} and ProvChain~\cite{Liang2017} are both blockchain-based solutions for data provenance accountability. Once collected, data provenance records are published to the blockchain network, verified, and eventually added to the blockchain. In other words, DataProv and ProvChain provide an immutable and secured ledger for data provenance records. Both solutions deal with data stored in the cloud. Even though these solutions seem to provide a way for researchers to comply with the GDPR’s right to access, how to comply with the rights to rectification and erasure is not addressed. Furthermore, compliance with licensing terms of shared data is not taken into account. 
Neisse et al.~\cite{Neisse2017} propose a blockchain-based solution for data provenance accountability. Their solution aims at empowering data subjects by enabling them to track who has accessed their data and whether these were used accordingly to their consent. It also aims at helping data controllers prove that they have received that consent, the proof of which is a smart contract to which both the data controller and subject are parties. The smart contract also encodes the policies for data access, usage, and transfer, as well as data provenance information. This solution focuses on the relationship between data subjects and data controllers. In contrast, the goal of LUCE is to define the generic interaction protocols between data providers (including data subjects) and data requesters.
The advantage of the LUCE approach is that data can be shared and reused by data requesters, while data providers and data subjects maintain control over the access and the type of reuse of data.
The Ocean Protocol~\cite{Ocean} is an industry-wide initiative to implement marketplaces for data sharing. This blockchain-based solution aims at facilitating the sharing of datasets (as well as algorithms and services, such as storage and processing) in a transparent, traceable, and trustworthy way, with the data providers keeping control over their datasets. However, the Ocean Protocol mainly targets companies that collect data. Indeed, data providers can choose to be monetarily rewarded for sharing their data and there is no considerations made for the data subjects. Even though the Ocean Protocol allows for traceability regarding what happens with datasets, it does not allow for the recording of the purpose for which the data have been used. Thus with OCEAN is not possible to guarantee compliance with the GDPR’s right to access, rectification and erasure. 
Finally, in \cite{shrestha2020blockchain}, a permissioned blockchain is adopted to support data sharing a GDPR compliant way. Compliance to GDPR is achieved by means of user consent. Consent can be revoked however the deletion of records is not explicitly addreseed. As in OCEAN, the authors define a compensation is required as a way to incentivize data-sharing. The solution differs to LUCE in several aspects, firstly the use of permissioned blockchain means that the decentralisation is limited by design in the platform. LUCE is envisioned as an open platform, engaging with end-users as data providers and data requesters. The way LUCE monitors data re-use is to include structured descriptions of purpose of use, as described in detail in \cite{jaiman2020consent}. Moreover, purpose of use and consent are dynamically checked and enforced within the smart contract.
\section{Conclusions and Future Work}
\label{sec:conclusions}
In this paper, we presented LUCE, a blockchain-based solution for automatic data management of licensing terms and accountability in a GDPR compliant manner. LUCE creates new opportunities for data sharing and reuse and makes it easier for researchers to track what happens to the dataset once they have been shared. Moreover, LUCE enables the enforcement of licensing terms and provides a solution for complying with the right to access, rectification, and erasure of GDPR. 
We show that with an increasing demand for data re-use LUCE scales similarly to any Ethereum based network in private and in public configurations of the network. The cost involved in the platform are greater for data providers. In follow up work we experimented with different incentive schemes \cite{jaimanincentive} and showed that the system can be calibrated to incentivise all the involved actors. Furthermore, this work acts as our base architecture model and we are continuously working on developing LUCE and its components:
\paragraph{Consent compliance} 
In the future, we will further investigate how to comply with the consent given by data subjects. The assumption here is that the license matches the consent of data subjects, which is true in most research datasets, however, there is a need for personalization of the individual consent of the data subjects which is also in line with existing legal frameworks. We already proposed an extension of LUCE~\cite{jaiman2020consent} which considers the consent codes developed by the Global Alliance for Genomic and Health~\cite{ConsentCodes, ga4gh} to systematically record data usage conditions based on the data subject's consent and purpose of use. This approach however is suitable for personal health data. We will investigate how to generalize the approach by identifying generic consent codes that support data collected in different domains. 
\paragraph{Data integration}
Another point that LUCE will address in the future is merging datasets for further reuse. It occurs in practice and is even one of the goals of data sharing and reuse. To this extent, we will seek the use of Digital Object Identifiers (DOI) in combination with a DOI suffix that tracks the merging of data to the original data sources. 
\paragraph{Evaluation} 
In this paper, we evaluate the technical feasibility and scalability of the LUCE platform. With LUCE, users can share data securely and transparently. Overcoming these two important barriers of data-sharing is expected to improve user engagement with data sharing practices and the benefits it brings at an individual and societal level. LUCE will need to be further evaluated on how well it helps users to perform data sharing and reuse. Therefore, behavioral evaluations are required to investigate to which extent the users of LUCE would be willing to make use of the platform and which features should be modified, removed, or added to ensure a broad user engagement.  
\section*{Acknowledgment}
This work was supported in part by the NWO Aspasia (Grant 91716421) and by the Maastricht York-Partnership Grant. The authors would like to thank Andine Haveange for her initial insights on the work which helped to define the LUCE architecture. They also would like to thank Prof. David Townend and Dr. Birgit Wouters for their insights on compliance with GDPR law and regulations. 

\bibliographystyle{unsrt}
\bibliography{bibliography}

\end{document}